\def\Re{{\rm Re}}
\def\Im{{\rm Im}}
\def\PiL{\Pi_{\rm L}}
\def\PiT{\Pi_{\rm T}}
\def\rhoT{\rho_{\rm T}}
\def\rhoL{\rho_{\rm L}}
\def\PL{P_{\rm L}}
\def\PT{P_{\rm T}}
\def\v{{\bf v}}
\def\A{{\bf A}}
\def\E{{\bf E}}
\def\B{{\bf B}}
\def\D{{\bf D}}

\def\j{{\bf j}}
\def\x{{\bf x}}
\def\p{{\bf p}}
\def\q{{\bf q}}

\def\bzeta{\mbox{\boldmath$\zeta$}}
\def\grad{\mbox{\boldmath$\nabla$}}
\def\half{{\textstyle{1\over2}}}
\def\CA{C_{\rm A}}
\def\eps{\epsilon}
\def\gammaE{\gamma_{\rm E}}

\def\MSbar{$\overline{\hbox{MS}}$}

\def\alphaw{\alpha_{\rm w}}
\def\Ns{N_{\rm s}}
\def\Nf{N_{\rm f}}
\def\bPhi{\mbox{\boldmath$\Phi$}}
\def\V{{\cal V}}
\def\c{{\bf c}}
\def\blambda{\mbox{\boldmath$\lambda$}}
\def\FAC{{\hbox{\tiny FAC}}}
\def\NLLO{{\hbox{\tiny NLLO}}}
\def\Const{{\cal C}}

\def\drangle{\rangle\!\rangle}
\def\dlangle{\langle\!\langle}
\def\bigdrangle{\bigr\rangle\!\bigr\rangle}
\def\bigdlangle{\bigl\langle\!\bigl\langle}
\def\deltaS{\delta^{S_2}}

\def\deltaC{\delta\hat C}

\documentstyle[epsf,preprint,aps,fixes]{revtex}
\tightenlines
\begin {document}


\preprint {UW/PT 99--25}

\title
{Non-perturbative dynamics of hot non-Abelian gauge fields:
beyond leading log approximation}

\author {Peter Arnold}

\address
    {%
    Department of Physics,
    University of Virginia,
    Charlottesville, VA 22901
    }%
\author{Laurence G. Yaffe}
\address
    {%
    Department of Physics,
    University of Washington,
    Seattle, Washington 98195
    }%
\date {December 1999}

\maketitle
\vskip -20pt

\begin {abstract}%
{%
Many aspects of high-temperature gauge theories,
such as the electroweak baryon number violation rate,
color conductivity, and the hard gluon damping rate,
have previously been understood only at leading logarithmic order
(that is, neglecting effects suppressed only by
an inverse logarithm of the gauge coupling).
We discuss how to systematically go beyond leading logarithmic order
in the analysis of physical quantities.
Specifically, we extend to next-to-leading-log order (NLLO)
the simple leading-log effective theory due to
B\"odeker that describes non-perturbative color physics in hot
non-Abelian plasmas.
A suitable scaling analysis is used to show that no new operators
enter the effective theory at next-to-leading-log order.
However, a NLLO calculation of the color conductivity is required,
and we report the resulting value.
Our NLLO result for the color conductivity can be trivially combined with
previous numerical work by G.~Moore to yield a NLLO result for the
hot electroweak baryon number violation rate.
}%
\end {abstract}

\thispagestyle{empty}


\section {Introduction}

The near equilibrium dynamics of hot, weakly-coupled non-Abelian plasmas,
such as high temperature QCD or electroweak theory, involves a surprisingly
rich hierarchy of spatial and temporal scales.
Of particular interest in recent years is the analysis of
non-perturbative processes occurring in (or near) equilibrium in such plasmas.
A primary application is to the rate of baryon number violation in
hot electroweak theory,
which quite possibly is responsible for the preponderance
of matter over anti-matter in our universe \cite{baryogenesis}.
Baryon number non-conservation is possible due to the electroweak anomaly
in the baryon (and lepton) number current,
and proceeds through non-perturbatively large thermal fluctuations in
electroweak gauge fields.
A key goal is to understand how to calculate the dynamics of such fluctuations.

As will be reviewed momentarily,
non-perturbative gauge dynamics in hot non-Abelian plasmas
may be described (at leading logarithmic order) by a remarkably
simple effective theory discovered by B\"odeker \cite{bodeker},
whose only input parameter is the value of the color%
\footnote{
  We use ``color'' as a generic term for some non-Abelian gauge charge,
  not as something specific to QCD.
}
analog of electrical conductivity.
The color conductivity in turn depends on what is known as the
hard gauge boson damping rate, which
is a measure of the rate at which collisions randomize a
charge carrier's color charge.
Up to now, however, these quantities have only been known
at leading order in the logarithm of the gauge coupling;
that is, neglecting effects suppressed only by $1/\ln(1/g)$,
where $g$ is the gauge coupling.
Consequently, numerical simulations of the high temperature
topological transition (or baryon number violation) rate
based on B\"odeker's effective theory 
have only been valid to leading order in logarithms \cite{moore}.
For practical purposes such leading log results are, by themselves,
not very useful --- there is a huge difference
between, say, $\ln(16 \pi/g)$ and $\ln(1/10 g)$ for any realistic value
of the gauge coupling.  But this difference is of sub-leading order in $\ln(1/g)$.
In this paper, we report a calculation of the color conductivity to
next-to-leading-log order (NLLO), and in the process explain what this
means in gauge-invariant language.
Our approach involves systematically constructing a sequence of effective
theories
which reach from short to long distance scales,
and performing perturbative calculations which enable one to match
the physics of each effective theory with its predecessor to the order desired.
When combined with Moore's numerical result \cite{moore}
for the constant of proportionality between the topological transition
rate and the color conductivity, our improved value for the
conductivity immediately yields a NLLO result for the topological transition rate.

In this paper, ``hot'' plasma means hot enough (1) to be ultra-relativistic,
(2) to ignore chemical potentials,
(3) for non-Abelian gauge couplings to be small, and (4) to be in the
high temperature symmetric phase.%
\footnote{
   We use the term ``symmetric phase'' loosely since, depending on the details
   of the Higgs sector, there need not be any sharp transition between the
   symmetric and ``symmetry-broken'' phases of the theory \cite{notransition}.
   A sharp transition is required for viable electroweak baryogenesis.
   Our analysis is applicable whenever the infrared dynamics of the
   Higgs field is irrelevant at lengths of $O(1/g^2T)$,
   which is the case either
   (a) far above the electroweak phase transition or ``crossover,'' or
   (b) in the symmetric phase at the transition in cases where there is
   a sufficiently strong first-order transition.
}
For convenient reference,
the critical distance and time scales associated with color dynamics
in such plasmas (to be reviewed below) are collected in
table \ref{tab:scales},
and the notation used in this paper is summarized in
table \ref{tab:notation}.


\begin {table}
\begin {center}
\begin {tabular}{l@{\quad}l}
\hline
$O(T^{\strut})$
    & typical momentum of excitations in the plasma. \\
$m = O(gT)$
    & inverse Debye screening length. \\
$\gamma = O[g^2T\ln(g^{-1})]$
    & hard gauge boson damping rate. \\
$O(g^2 T)$
    & inverse spatial scale of non-perturbative gauge fluctuations. \\
$O[g^4 T\ln(g^{-1})_{\strut}]$
    & inverse time scale of non-perturbative gauge fluctuations. \\
\hline
\end {tabular}
\end {center}
\caption
   {%
   \label {tab:scales}
   Important (inverse) distance and time scales.
   }%
\end {table}


\begin {table}
\begin {center}
\begin {tabular}{l}
\hline
$v^\mu = (1,\v)$; $\v$ a spatial unit vector. \\
$\A = \A(\x,t)$, the spatial non-Abelian gauge field. \\
$\D = \grad + g\A^a T^a$, the gauge covariant derivative. \\
$W = W(\x,\v,t)$, the adjoint color distribution of hard excitations. \\
$\bzeta = \bzeta(\x,t)$ and $\xi = \xi(\x,\v,t)$ are Gaussian white noise. \\
$\dlangle \cdots \drangle$ denotes averaging over noise. \\
$\langle \cdots \rangle \equiv \langle \cdots \rangle_\v$ denotes
   averaging over the direction $\v$. \\
$\deltaS(\v{-}\v')$ is a $\delta$-function on the two-sphere
    normalized so that $\langle \deltaS(\v{-}\v')\rangle_{\v'} = 1$. \\
$\deltaC \, W \equiv \langle \delta C(\v,\v')\, W(\v') \rangle_{\v'}$,
    the linearized collision operator applied to $W$. \\
$\CA$ is the adjoint Casimir of the gauge group [$N$ for SU($N$)]. \\
$d = 3-\eps$ with $\eps\to0$, the number of spatial dimensions. \\
\hline
\end {tabular}
\end {center}
\caption
   {%
   \label {tab:notation}
   Summary of notation.
   }%
\end {table}


\section {Review}

It is well known that non-perturbative gauge field fluctuations
in hot non-Abelian plasmas
are essentially magnetic and characterized by the
distance scale%
\footnote{
   See the introduction of Ref.~\cite{Blog1} for a simple
   heuristic argument, or Ref.\ \cite{scale} for more formal ones.
}
$R \sim (g^2 T)^{-1}$.
The corresponding time scale has only
been understood more recently, and can be arrived at by the following
simple physical argument.  Start with (non-Abelian) Ampere's Law,
$\D\times\B = D_t \, \E + \j$.  Now recall that plasmas are conductors,
and so write $\j = \sigma \E$, where $\sigma$ is the (color)
conductivity.
If we assume the time dependence will be slow and treat the covariant time
derivative $D_t \, \E$ as negligible compared to $\sigma \E$
(an approximation that can be verified {\it a posteriori}), then
the result is
\begin {equation}
   \D\times\B = \sigma \E \,.
\end {equation}
Extracting the time scale from this equation is clearest if one specializes
to $A_0 = 0$ gauge:
\begin {equation}
   \D\times\D\times\A = \sigma {d\over dt} \A \,.
\label {eq:ampere1}
\end {equation}
In terms of characteristic scales, this is just
\begin {equation}
   R^{-2} \A \sim \sigma \, t^{-1} \A ,
\label {eq:R2simt}
\end {equation}
and so the characteristic time $t \sim \sigma R^2 \sim \sigma/g^4 T^2$.
The color conductivity $\sigma$ was originally analyzed by Selikhov and
Gyulassy \cite{Selikhov&Gyulassy} and is order
$\sigma \sim T/\ln(1/g)$, yielding
$t \sim [g^4 T \ln(1/g)]^{-1}$.
The rate per unit volume $\Gamma$ for topological transitions in a
hot, non-Abelian plasma must therefore have the characteristic size
$\Gamma \sim 1/(R^3 t) \sim \alpha^5 T^4 \ln (1/\alpha)$,
where as usual $\alpha \equiv g^2 / 4\pi$.
The factor of $\alpha^5$ in this rate was originally noted by ourselves and
Son \cite{alpha5}, and the presence of the logarithm by B\"odeker
\cite{bodeker}.
The explanation in terms of color conductivity is given in detail in
Ref.\ \cite{Blog1}.

The above order of magnitude arguments can be made quantitative.
One can numerically integrate, with Boltzmann-weighted initial conditions,
the classical equation (\ref{eq:ampere1}) that we derived from Ampere's Law,
and measure the frequency of topological transitions \cite {moore}.
A classical, as opposed to quantum, treatment is adequate
because the low momentum modes ($k \ll T)$ of the gauge field in a
hot plasma have large occupation numbers
[$n_k \sim {(e^{k/T}-1)^{-1}} \sim T/k \gg 1$] and so, by the
correspondence principle, behave classically \cite{classical}.
Corrections to the classical approximation are suppressed by
powers of $k/T$.
In our case, we are interested in modes with $k \sim R^{-1} \sim g^2 T$,
which is indeed small compared to $T$ for weak coupling.

However, Eq.\ (\ref{eq:ampere1})
first requires some refinement.  It is a purely
dissipative equation.  An effective theory for the equilibrium dynamics
of the system must therefore incorporate a source of thermal noise to
maintain thermal equilibrium.  Ultimately, this noise is (like the
dissipation) due to interactions of the modes of interest with other
degrees of freedom of the system, specifically typical charge carriers, which
have momentum of order $T$.  As discussed in Ref.\ \cite{Blog1},
one may argue from general principles that, at the distance and time
scales of interest, this noise can be taken to be simple Gaussian
white noise, appropriately normalized to produce the desired
temperature $T$.  The resulting effective theory is%
\footnote{
    This equation certainly makes sense in $A_0=0$ gauge,
    and is also correct and unambiguous in
    general flow gauges of the form $A_0 = R[\A]$,
    where $R[\A]$
    depends on $\A(\x,t)$ only
    instantaneously and so does not involve time derivatives of $\A$.
    See Ref.\ \cite{flow gauges} for a discussion of flow gauges,
    and Ref.\ \cite{zinnjustin&zwanziger} for a proof that
    the equation (\ref{eq:bodeker}) may be applied in any gauge of this class.
    There are subtleties
    in directly interpreting the Langevin equation (\ref{eq:bodeker})
    in {\it other}\/
    gauge choices, such as Landau gauge.
    For various historical reasons, our approach for doing
    actual calculations
    is to use a path integral representation of these Langevin equations,
    for which it is straightforward to fix any gauge desired using
    the usual Faddeev-Popov procedure.  (See Ref. \cite{thy2} for details.)
    We find it convenient, in particular, to use Coulomb gauge.
    But Coulomb gauge might also be implemented as the
    $\lambda \to \infty$ limit of the flow gauge $A_0 = \lambda \grad\cdot\A$,
    for which there should be no problem working
    directly with the Langevin formulation (\ref{eq:bodeker}) instead of
    the associated path integral representation.
}
\begin {mathletters}
\label {eq:bodeker}
\begin {equation}
   \D\times\B + \bzeta = \sigma \, \E \,,
\end {equation}
where $\bzeta$ is Gaussian noise with correlation
\begin {equation}
   \bigdlangle \zeta_i(t,\x) \zeta_j(t',\x') \bigdrangle
   = 2 \sigma T \, \delta_{ij} \, \delta(t{-}t') \, \delta^{(3)}(\x{-}\x') .
\end {equation}
\end {mathletters}%
The effective theory (\ref{eq:bodeker}) is also known as
stochastic 3-dimensional gauge theory, and has a long history of theoretical
study motivated by formal questions divorced from any connection with
high temperature dynamics \cite{stochastic}.
A key property of this theory is the fact that it is UV finite \cite {Blog1}.

The final ingredient needed to make the effective theory quantitative,
is the precise value of the color conductivity $\sigma$.
This has previously been known only at leading log order 
\cite{Selikhov&Gyulassy,bodeker,Blog1}:
\begin {equation}
   \sigma \approx  {m^2 \over 3\gamma},
\label {eq:sigmaLLO}
\end {equation}
where $\approx$ denotes equality up to relative corrections suppressed by
$[\ln(1/g)]^{-1}$.
Here, $m$ is the leading-order Debye screening mass,
which is well known,%
\footnote{
   For SU($N$) gauge theory with $\Ns$ scalars and $\Nf$ Dirac fermions in the
   fundamental representation,
   the Debye mass
   $m^2 = \frac 16 (2 N + \Ns + \Nf) \, g^2 T^2$
   %
   %
   at leading order in $g$.
   For hot electroweak theory with three fermion families and
   a single Higgs doublet,
   $m^2 = {11\over6} \, g^2 T^2$.
}
is $O(gT)$, and
completely encapsulates the dependence on the matter field content
of the underlying theory.
$\gamma$ is the hard gauge boson damping rate \cite{gammag} and,
to leading-log order, is given by
\begin {equation}
   \gamma \approx \CA \alpha T \ln\!\left(1\over g\right) .
\label {eq:gamma0}
\end {equation}
The leading-log effective theory represented by (\ref{eq:bodeker}) and
(\ref{eq:sigmaLLO}) was first derived by B\"odeker \cite{bodeker}.
(For subsequent derivations,
see Refs.\ \cite{Blog1,moore,manuel,basagoiti,b&i}.%
\footnote{
   The derivation in Ref.\ \cite{manuel} treats the color charge of the charge
   carriers as classical, using Wong equations.
   This approach has been used before in
   analyzing hot gauge theories, and it is worth pointing out one way to
   understand, after the fact, why it works.  In the case at hand, note
   that the result (\ref{eq:sigmaLLO}) for $\sigma$ does not depend on the
   color representations of the charge carriers in any way except
   implicitly in the value of $m^2$.  At leading order, the contribution
   to $m^2$ of any charge carrier species is proportional to the adjoint
   Casimir $C_R$ for the color representation of that species, and there is
   no other representation dependence.  If one extracts the proportionality
   constant for representations with arbitrarily large values of
   $C_R$, one
   will then also automatically have the right proportionality constant for
   small representations.  But the color of large representations can be
   treated classically.
   So treating the color of the charge carriers classically and then
   setting the square of the classical charge to
   the quantum value $Q^2 = C_R$ at the end of the day works because
   of the simplicity of the color structure for the quantities of interest.
})
Using this theory, Moore \cite{moore}
has numerically simulated the topological transition
rate for electroweak theory, obtaining
\begin {equation}
   \Gamma \approx (10.8 \pm 0.7)
       \left( {2\pi \over 3\sigma} \right)
       (\alpha T)^5 \,.
\label{eq:GammaB}
\end {equation}


\section {The sequence of effective theories}

The series of effective theories we will need to describe color
dynamics at low frequency $\omega$ and momentum $k$
[ultimately $\omega \sim t^{-1} \sim [g^4 T \ln(1/g)]^{-1}$
and $k \sim R^{-1} \sim (g^2 T)^{-1}$] are as follows.

\begin {mathletters}
\label {eq:1}
\begin {eqnarray}
\noalign {\hbox{\it Theory 1: $\omega,k \ll T$}}
&
   (D_t + \v\cdot\D)\, W - \v\cdot\E = 0 \,,
\label {eq:1a}
&\\&
   D_\nu F^{\mu\nu} = \jmath^\mu = m^2 \langle v^\mu W \rangle \,.
&
\end {eqnarray}
\end {mathletters}

\begin {mathletters}
\label {eq:2}
\begin {eqnarray}
\noalign {\hbox{\it Theory 2: $\omega \ll k \ll m$}}
&
   \v\cdot\D \, W - \v\cdot\E = -\deltaC \, W + \xi \,,
\label {eq:2a}
&\\&
   \langle W \rangle = 0 \,,
\label {eq:2b}
&\\&
   \D \times \B = \j  = m^2 \langle \v W \rangle \,,
&\\&
   \dlangle \xi \xi \drangle =
  {\displaystyle{2 T\over m^2} \, \deltaC} \,.
&
\end {eqnarray}
\end {mathletters}

\begin {mathletters}
\label {eq:3}
\begin {eqnarray}
\noalign {\hbox{\it Theory 3: $\omega \ll k \ll \gamma$}}
&
   \sigma \E = \D \times \B + \bzeta \,,
&\\&
   \dlangle \bzeta \bzeta \drangle
	  = 2 \sigma T \,.
&
\end {eqnarray}
\end {mathletters}
All theories have been written in forms that are local in
space and time.
All fields should be interpreted as classical.

\medskip

{\it Theory 1.}
The first effective theory is a now-standard formulation
of ``hard-thermal-loop'' dynamics \cite {HTL} which is valid
for $\omega,k \ll T$, up to corrections suppressed by powers of $g$.
These equations amount to linearized, collisionless, non-Abelian,
Boltzmann-Vlasov kinetic theory.
One has conceptually split the degrees of freedom in the underlying
quantum field theory into those associated with quanta that have
large momenta $p \sim T$ and those with low momenta $k \ll T$.
The high momentum (or ``hard'') quanta are treated collectively
by a linearized Boltzmann equation;
the (bosonic) low momenta modes, referred to as ``soft''
(and sometimes also semi-hard) modes are,
because of the high occupation numbers mentioned earlier,
treated as comprising a classical field.
The first equation (\ref{eq:1a}) is a collisionless linearized
Boltzmann
equation for the propagation of the hard quanta in the presence of the
soft gauge fields.  $W(\x,\v,t)$ represents the color distribution of
hard particles in space, velocity, and time, where the velocity $\v$
is a unit vector because all hard particles are ultra-relativistic,
and where $\langle \cdots \rangle$ in Eqs.\ (\ref{eq:1}) and
(\ref{eq:2}) denotes averaging over the direction of $\v$.
Although the hard particles have individual momenta $p \sim T$ associated
with very short wavelengths, the collective distribution of hard
particles, described by $W$, can have slow spatial variation
(in other words, the density of hard particles may vary slowly over
a large region of space), which is why $W$ appears in the
effective theory describing $\omega, k \ll T$ dynamics.

The distribution $W$ appearing in the above effective theories
only encodes information about the color structure of
collective fluctuations, and $W$ lives in the adjoint representation
of the gauge group.%
\footnote{
    Technically,
    $W$ is the adjoint representation piece of the density matrix
    describing the color charges of the hard excitations,
    summed over the various species of excitations and
    integrated over the energy of excitations
    (for a fixed direction of motion $\v$).
    It is normalized in a way that simplifies the resulting equation.
}
(See Refs.\ \cite{Blog1,bodeker,iancu} for details.)
A full description of all physics for $\omega, k \ll T$ would also contain
additional distribution functions describing fluctuations
that do not contribute to the color current.
These include color-neutral fluctuations which are responsible
for hydrodynamic phenomena such as viscosity and sound,
but which do not couple to the long-distance color
dynamics except at higher order in the coupling $g$.
We will therefore ignore them.%
\footnote{
   There are also other colored sectors besides the adjoint
   one represented by $W$.  
   For each species of hard excitations,
   there are sectors corresponding to every irreducible representation
   contained in $R\times\bar R$, where $R$ is the color representation
   of the given excitation.
   See Ref.\ \cite{Blog1} for a discussion.
   However, we
   do not know of any interesting physics that
   specifically couples to the other colored
   sectors except at sub-leading order in $g$.
}
For a discussion of a tower of effective theories relevant to
the uncolored sector, see Ref.\ \cite{jj}.

There are several things we have left out of Theory 1 because
they do not affect color dynamics at leading order.
One is a collision term in the Boltzmann equation (\ref{eq:1a}) due to
hard collisions.  The cross-section for such collisions is $O(\alpha^2)$
and is overwhelmed, for color dynamics, by softer collisions that
will be discussed in the context of Theory 2 below.%
\footnote{
   Hard collisions are relevant at leading order in $g$ to physics in
   the uncolored sector, however.  See, for example, Ref.\ \cite {viscosity}.
}
We have also left out the soft modes of non-gauge fields such as
scalars or fermions.  These do not affect the color conductivity or
the final soft dynamics (Theory 3) at leading order in $g$,
provided the effective thermal mass of any colored scalars is large enough
[large compared to $g^2 T$]
so that they decouple at the very soft scales of interest for Theory 3.

\medskip

{\it Theory 2.}
The $\omega \ll k \ll m$ limit of Theory 1 is a small frequency version
of a theory originally written down by B\"odeker \cite{bodeker},
and is discussed in
detail in Ref.\ \cite{thy2}.
In this low frequency regime, all time derivative terms in Theory 1
become negligible, and have been dropped, except for the
one implicit in $\E$ (which is $-d\A/dt$ in $A_0=0$ gauge).
Gauss' Law is replaced by the constraint (\ref{eq:2b})
due to the effects of Debye screening for $k \ll m$.
(See Ref.\ \cite {thy2} for details and interpretation.)

The most physically significant change is the introduction of the collision
term $\deltaC \, W$ into the linearized Boltzmann equation (\ref{eq:2a}).
This term represents the effects of $2 \to 2$ collisions of hard particles via
the $t$-channel exchange of what are called semi-hard gauge bosons
($\omega \lesssim k \lesssim m$).  The leading log calculation of the
linearized collision operator $\deltaC$
was first made by B\"odeker \cite{bodeker} and is explained in the language of
collisions in Ref.\ \cite{Blog1}.
One finds,
\begin {mathletters}
\label {eq:deltaC}
\begin {equation}
   \deltaC \, f(\v) \equiv \langle \delta C(\v,\v') \, f(\v') \rangle_{\v'} ,
\end {equation}
\begin {equation}
   \delta C(\v,\v') \approx
    \gamma(\mu) \left[
       \deltaS(\v{-}\v')
       - {4\over\pi} \> {(\v\cdot\v')^2 \over \sqrt{1-(\v\cdot\v')^2}} 
    \right] ,
\label {eq:delC-LL}
\end {equation}
\end {mathletters}
where
\begin {equation}
 \gamma(\mu) \approx \CA \alpha T \ln\!\left(m \over \mu\right)
\end {equation}
is the leading log contribution to the hard gauge boson
damping rate due to exchange of gauge bosons having spatial momentum $q > \mu$.
In other words, in a Wilsonian view of renormalization, $\mu$ is the
ultraviolet cut-off imposed on $k$ to define effective theory 2, and
$\gamma(\mu)$ is the damping rate due to gauge field
fluctuations which have been integrated out and are no longer present
in the effective theory.
(In practice, it will be more convenient to perform renormalization by
standard subtraction methods, and $\mu$ will be the renormalization scale
rather than an ultraviolet cut-off.)
The normalization of the angular delta function $\deltaS(\v)$ is given in
table \ref{tab:notation}.

The collision term in Theory 2
causes dissipation in the effective theory and so
an appropriately matched thermal noise term must also be present.
The $\xi$ in (\ref{eq:2a}) is Gaussian noise with correlation \cite{bodeker}
\begin {equation}
      \dlangle \xi(\v,\x,t) \, \xi(\v',\x',t') \drangle
      = {2 T\over m^2} \, \delta C(\v,\v')
        \, \delta(t{-}t') \, \delta^{(3)}(\x{-}\x') .
\end {equation}
A path integral representation of this effective theory may be found
in Ref.\ \cite{thy2}.

\medskip

{\it Theory 3.}
This is B\"odeker's effective theory, discussed earlier, and is
obtained at leading log order by dropping $\mu$ below $\gamma$
(the momentum scale below which Theory 3 becomes valid) in Theory 2,
and then solving the $W$ equations for
$k \ll \gamma$
[which allows one to drop $\v\cdot\D$ compared to $\deltaC$ in (\ref{eq:2a})].
Details may be found in Refs.\ \cite{bodeker,Blog1}.
We emphasize that, for the purpose of expanding in inverse powers
of logarithms, we formally consider the momentum scale $k \sim g^2 T$
of ultimate interest to be parametrically small compared to
$\gamma \sim g^2 T \ln(g^{-1})$.

The decoupling of massive fields from low energy physics is very
familiar in Euclidean field theory,
and forms the basis for traditional applications of effective field
theory techniques.
In the case at hand, it is possible to eliminate the field $W$
(while maintaining a local description in the effective theory)
because one is dropping below the scale $\gamma$
characterizing the decay time of color correlations.
This harks back to an old suggestion by Lebedev and Smilga
\cite{Lebedev&Smilga} that $\gamma$ might cut off infrared divergences
for some quantities.
(See also Ref.~\cite {USA}.)
As we shall see, this indeed happens for the color conductivity at
next-to-leading log order, in a sense that we will make precise.

It has long been known that the leading-order hard-thermal-loop effective
interactions between gauge fields with momentum $k \sim m \sim gT$ are
non-local in space and time.
That is, formally eliminating $W$ from the Theory 1 Eqs.\
(\ref{eq:1}) generates non-local interactions for $\A$.
It should be emphasized that it is the inclusion of collisions
which allows one to recover locality of effective gauge interactions
for $k \ll \gamma$, as in Theory 3.
In the non-Abelian theory, arbitrarily small angle collisions will
cause the color correlations of hard particles to
fall off rapidly for times and distances large
compared to the relevant collision time $1/\gamma$,
permitting one to treat
such correlations as local
in an effective theory restricted to $k \ll \gamma$.
Leading-order calculations using hard-thermal-loop
interactions miss this effect because
they
do not incorporate the effect of collisions
on the propagation of hard color fluctuations.
In such a collisionless approximation, the disturbance created by
probing the hard particles at one time
does not decay, and can have an arbitrarily long-lasting effect on
fluctuations measured at a later time---hence, non-locality.
With collisions included, this non-locality is cut-off at the scale
$1/\gamma$.


\section {Scaling and subleading corrections}


\subsection {Review of static case}

One of the great advantages to using effective theories is that,
besides simplifying the description of the relevant physics at a given
scale, they provide a very clean mechanism for organizing, analyzing,
and computing sub-leading corrections.
For hot gauge theories, the quintessential example is the
analysis of {\it static}\/ 
equilibrium properties
by Braaten and Nieto \cite{Braaten&Nieto}
using a sequence of Euclidean effective theories.
(See also Ref.\ \cite{KLRS}.)
The important scales for static equilibrium physics are $T$, $gT$, and $g^2 T$;
the scale $\gamma$, which will be crucial to our discussion of dynamics,
does not appear.  Below $T$, the effective theory is a three-dimensional
Euclidean gauge theory coupled to an adjoint scalar
$A_0$ with mass $m = O(gT)$, this mass being the manifestation of Debye
screening.  Below the scale $m$, the $A_0$ field decouples,
and one is left with a three-dimensional theory of unscreened
magnetic physics:
\begin {equation}
   S_{\rm eff} = T^{-1} \int d^3 \x \left[
        \half F_{ij}^a F_{ij}^a + \cdots
   \right] .
\label{eq:3d}
\end {equation}
The term shown explicitly is simply renormalizable (in fact finite)
3-dimensional pure gauge theory.  The dots represent an infinite
sequence of interactions with higher and higher scaling dimension
($F^3$, $F^4$, {\em etc}).
These terms are infrared irrelevant, in the sense of
the renormalization group, and their importance is suppressed
by higher and higher powers of the scale $k$ of the physics of interest,
compared to the scale $\Lambda = m \sim gT$ where
the effective theory breaks down.
So, for the study of non-perturbative physics, where $k \sim g^2 T$,
terms in $S_{\rm eff}$ with higher and higher scaling dimension
correspond to corrections whose effects are suppressed by
higher and higher powers of the ratio $k/m \sim g$.

It is important to note that the coefficients of specific higher order terms
may also contain explicit factors of $g$, which will cause their effects
to be even more suppressed than indicated by their scaling dimension alone.

Standard power counting of 3-dimensional gauge theory (\ref{eq:3d})
gives the ultraviolet scaling dimension of the field $\A$ to be
$[A] = [x]^{-1/2}$ (or equivalently $[k]^{1/2}$).
One typically rescales the field
$\A \to T^{-1/2}\A$ so that the coefficient of the kinetic
term in (\ref{eq:3d}) is dimensionless and so that counting scaling
dimension is then the same as counting engineering dimension.
In this paper, however, we shall avoid such redefinitions in order
not to obscure the relationship between the gauge fields appearing in
successive effective theories.

Braaten and Nieto showed how, in principle, one can
calculate the coefficients of the higher-dimensional terms in
the effective theory (\ref{eq:3d}), to whatever order desired,
by carrying out perturbative matching calculations.
Perturbative calculations suffice because
the effective theory is valid for scales $k \ll m$,
and because the physics of gauge field fluctuations is
perturbative at the upper end $g^2 T \ll k \ll m$ of that
range of validity.
Roughly speaking, the idea is to pick a set of physical quantities that probe
physics at a scale $k \ll \Lambda$ where the effective theory is valid,
then to calculate those quantities in both the effective theory and
the underlying theory, and then fix the (finite) set of unknown
coefficients (at a given order) by requiring the answers to be the same.
A refinement is required to make the calculation tractable,
because long-distance
physics in the effective theory (\ref{eq:3d}) [and so also in the
underlying theory] is non-perturbative!
Conceptually, one imagines restricting the theory to a large but not
too large box --- one of size $L$ which is large enough
($L \gg \Lambda^{-1}$) for the
effective theory expansion to be valid, but small enough ($L \ll (g^2T)^{-1}$)
that all the physics in the box is perturbative.  The matching calculations can
then be performed perturbatively.

The ``large box'' temporarily introduced for the sake of matching
can be replaced by any suitable infrared regularization,
and in practice it is most convenient to use dimensional regularization.
Dimensional regularization is also used for
ultraviolet regularization, if required to define the theory
(at a given order) in the first place.


\subsection {The dynamic case}

Given that our goal is to expand in inverse powers of logarithms
$\ln(g^{-1})$, while always working to leading order in powers of $g$,
trying to refine our first two effective theories (\ref{eq:1}) and (\ref{eq:2})
by introducing new infrared-irrelevant interactions is unnecessary.
We need to be able to use Theory 1 at the scale $m$ in order to match onto
Theory
2, but then irrelevant interactions are suppressed by powers of
$m/T \sim g$, not by mere powers of inverse logs.
We need to be able to use Theory 2 at the scale $\gamma$ in order
to match onto Theory 3, but then irrelevant interactions are suppressed
by powers of $\gamma/m \sim g\,\ln(g^{-1})$, which can again be ignored.
The crux of an expansion in inverse logs, as far as the introduction of
higher-dimensional interactions is concerned, is in the application of Theory 3
(B\"odeker's effective theory)
to non-perturbative physics at a scale $k \sim g^2 T \ll \gamma$.
The influence of higher dimension irrelevant terms will then only be
suppressed by factors of $k/\gamma \sim 1/\ln(g^{-1})$.

Fortunately, a delightful simplification occurs, to be explained momentarily:
effects of adding any
higher-dimension corrections to B\"odeker's effective theory
(\ref{eq:3}) are suppressed by {\it more}\/ than one power
of the inverse logarithm.
This means that B\"odeker's effective theory is already perfectly
adequate at next-to-leading log order, provided one calculates its one
parameter (namely $\sigma$) to next-to-leading log order.
Previous numerical simulations of B\"odeker's theory \cite{moore}
can therefore be instantly extended to NLLO accuracy simply by using
the improved value of $\sigma$.

Here is one version of the power counting analysis which demonstrates
the required suppression of higher-dimension operators.
It is possible to turn the Langevin equation (\ref{eq:3}) that defines
the effective theory into a supersymmetric path integral form
that allows standard renormalization analysis \cite{thy2}, which we review
in the appendix.  But one can get to the same result in a simpler, more
cavalier way.
The dynamical effective theory has to (and does) produce the same
equilibrium distribution that is more traditionally analyzed with
the Euclidean effective theories discussed earlier.
We can therefore borrow the well-known result that the field
$\A$ scales with distance as $[A] = [x]^{-1/2}$ in the ultraviolet.
That means that, at short distance, $\D \to \grad$ and B\"odeker's
effective theory becomes the free theory
$\sigma \E = \grad \times \B + \bzeta$.
This is just $-\sigma \, d\A/dt = \grad \times \grad \A + \bzeta$
in $A_0 = 0$ gauge,
and implies that time scales as $[t] = [x]^2$.
This, of course, is just a re-iteration of (\ref{eq:R2simt}).
The lowest dimension terms
that could possibly be added to the equation (\ref{eq:3}) for the effective
theory, consistent with parity and gauge-invariance,
are $\E\times\B$ and $\B\times\D\times\B$.
These terms have scaling dimension $[x]^{-4}$ as opposed to
those in B\"odeker's equation (\ref{eq:3}), which are $[x]^{-5/2}$.
The difference of 3/2 in scaling dimension means their effect is
suppressed%
\footnote{
   It should be noted that this power counting argument gives the
   correct suppression factor when comparing terms
   that are either infrared irrelevant or marginal, and which
   do not have unnaturally small coefficients at the scale of the cutoff
   on the effective theory.
   It is the derivative terms in the uncorrected effective theory
   which define marginal scaling.
}
by $(k/\gamma)^{3/2}$.
So, based simply on scaling dimension alone, the effect of possible higher
dimension operators, when $k = O(g^2 T)$,
is suppressed by at least $[\ln(g^{-1})]^{-3/2}$.
This substantiates our claim that the form of B\"odeker's equation
(\ref {eq:3}) remains unchanged to next-to-leading log order.
The effects of higher dimension terms may, of course, be even smaller
that this estimate if there is explicit suppression in their coefficients.
In fact, as discussed in the appendix,
the coefficients of these operators necessarily include a factor of $g$
which implies a further suppression of $(g^2 T/\gamma)^{1/2}$.
Hence, at the $g^2 T$ scale, the actual relative correction due to
higher-dimension operators is at least $[\ln(g^{-1})]^{-2}$.


\subsection {Implications for the meaning of \boldmath$\sigma$}

The above discussion allows us to provide an unambiguous definition of
what we {\it mean}\/ by the color conductivity at NLLO: we simply define
it to be the coefficient of B\"odeker's effective theory (\ref{eq:3})
at that order.
This is a non-trivial issue because the conventional definition
of conductivity
for an Abelian gauge theory,
in terms of the linear response of the current $\j$ to a
static, homogeneous electric field $\E$,
does not generalize in a meaningful way to a non-Abelian theory.
This is because, in a non-Abelian theory,
there is no gauge-invariant way to define
the induced current $\j$, nor is there any straightforward way to
define the imposition of a ``homogeneous'' background
electric field on top of the thermally fluctuating gauge field.%
\footnote
    {
    Actually,
    one can give a sensible meaning to a uniform non-Abelian gauge field
    \cite {constant-field}.
    But this does not generalize in a natural way to defining the
    addition of a uniform external field on top the thermal fluctuations
    in the gauge field.
    }
In contrast, B\"odeker's effective theory is a local theory,
and Eq.\ (\ref{eq:3}) is nicely gauge covariant with $\sigma$
a gauge-invariant numerical constant.%
\footnote{
  The alert reader may object that we earlier expressed uncertainty
  as to whether Eq.~(\ref{eq:3}) is correct (and unambiguous) in gauges
  other than flow gauges.
  However, the Langevin equation (\ref{eq:3}) can be reformulated
  as a manifestly gauge-invariant path integral, as discussed in
  Ref.\ \cite {thy2}.
}

Beyond NLLO, however, the definition of $\sigma$ as the parameter of the
effective theory presumably becomes ambiguous.  That's because at
next-to-next-to-leading log order (NNLLO), we need to introduce
irrelevant infrared interactions into the theory, namely
$\E \times \B$ and $\B \times \D \times \B$, as discussed earlier.
Such interactions render the theory non-renormalizable and so require
the introduction of UV regularization to what was previously a UV-finite
effective theory.
That is entirely normal for an effective theory, but it means that
the parameters of the theory will now be convention dependent,
depending on the arbitrary choices of renormalization scheme and
renormalization scale.  So, barring a conspiracy, we only have a clean,
unambiguous definition of $\sigma$ at NLLO.
It is entirely possible that color conductivity is only an approximate
concept and that there is no natural, unambiguous definition beyond
this order.

An unambiguously defined conductivity is, of course, irrelevant to the success
of the effective theory at any order, just as (the lack of) a
convention-independent
coupling constant is irrelevant to the applicability of
familiar zero-temperature perturbation theory.


\section {The result for \boldmath$\sigma$}

The details of our calculation of NLLO color conductivity are given in
Ref.\ \cite {sigma}, but we will summarize the main results here.
Throughout, we use dimensional regularization
in $d=3-\eps$ spatial dimensions with gauge coupling
$\mu^{\eps/2} g$.


\subsection {Matching Theory 1 to Theory 2}

Matching Theory 1 to Theory 2 involves calculating,
to the requisite accuracy,
the linearized collision operator $\deltaC$
which results from the effects of gauge field fluctuations in Theory 1,
in the presence of a perturbative infrared cut-off.
Conceptually, one must then repeat this calculation in Theory 2
taking into account both the bare collision operator in Theory 2,
and the residual effects of gauge field fluctuations which remain
in Theory 2.
A major virtue of using dimensional continuation for both infrared
and ultraviolet regularization is that,
with this choice of regulator,
the total collision operator in Theory 2 equals the bare operator;
the residual gauge field fluctuations in Theory 2 have no net effect.

The computation in Theory 1 may be carried out using any of the approaches
that have been used for leading log results \cite{bodeker,Blog1,moore,manuel}.
The kernel of the collision operator is given by
\begin {equation}
   \delta C(\v,\v') =
   {\CA m^2 T \over 2 g^2 \mu^\eps} \left[
          \left\langle \int_\q |{\cal M}(\v,\v',\q)|^2 \right\rangle_{\v'}
              \deltaS(\v{-}\v')
          - \int_\q |{\cal M}(\v,\v',\q)|^2
   \right] ,
\label {eq:dc2}
\end {equation}
with the scattering amplitude ${\cal M}$ which results from $t$-channel gauge boson
exchange with hard-thermal-loop (HTL) self-energies included,
\begin {equation}
   \int_\q |{\cal M}|^2
    = g^4 \mu^{2\eps} \int {d^{4-\eps} Q\over (2\pi)^{4-\eps}} \>
      \left| v_\mu \left(
          {\PT^{\mu\nu}(Q)\over Q^2 + \PiT(Q)} +
          {\PL^{\mu\nu}(Q)\over Q^2 + \PiL(Q)}
      \right) v'_\nu \right|^2
      \,2\pi \delta(Q\cdot v) \> 2\pi \delta(Q\cdot v') \,.
\label{eq:Msqr1}
\end {equation}
($\PT$ and $\PL$ are transverse and longitudinal projection operators,
respectively.
See Ref.~\cite {sigma} for details.)
The resulting form of the kernel $\delta C(\v,\v')$, accurate to NLLO, is
much more
complicated that the leading-log expression (\ref {eq:delC-LL}).
Fortunately, the only portion of $\delta C$ which is actually needed
for the determination of the NLLO color conductivity is the single matrix
element
\begin {equation}
    \gamma_1 \equiv \langle v^i \, \deltaC \, v^i \rangle_\v \,.
\end {equation}
At leading-log order, this is the same as the hard gauge boson damping rate
(\ref {eq:gamma0}), but at NLLO these quantities differ.
We find \cite {sigma} that 
to leading order in $g$ and all orders in $1/\ln(g^{-1})$,
\begin {equation}
   \gamma_1(\mu) =
   \CA \alpha T \left[
       - {1\over\eps}
       + \ln\!\left(m\over\bar\mu\right)
       + \half \ln {\pi \over 4}
       + a_1
   \right] ,
\label{eq:gamma}
\end {equation}
where we have written the result in terms of the \MSbar\ scale
$
   \bar\mu = \sqrt{4\pi} e^{-(\gammaE/2)} \mu \,.
$
The constant $a_1$ is given by the integral
\begin {eqnarray}
   a_1 \equiv
   {1 \over 4}
   \int_{-1}^{+1} d\lambda \>
   \Biggl\{ &&
          (1{-}\lambda^2)
          \left[
              {1 \over 2}\,
             {\arg \rhoT(\lambda) \over \Im \, \rhoT(\lambda)}
              + {\arg \rhoL(\lambda) \over \Im \, \rhoL(\lambda)}
	      + \Re \, {\ln \rhoT^*(\lambda) - \ln\rhoL(\lambda) \over
                  \rhoT^*(\lambda) - \rhoL(\lambda)}
	  \right]
          - {1\over |\lambda|}
    \Biggr\} \,.
\label{eq:aeta}
\end {eqnarray}
with
\begin {mathletters}
\label {eq:rho}
\begin {eqnarray}
   \rhoL(\lambda) &=& 1 - {\lambda\over2} \ln\!\left( 
                                   1+\lambda\over1-\lambda \right)
                      + {i\pi\over2} \, \lambda \,,
\\
   \rhoT(\lambda) &=& {1\over2} \left[{1\over1-\lambda^2}
                                       - \rhoL(\lambda)\right] .
\end {eqnarray}
\end {mathletters}%
(A more explicit integral representation for $a_1$ may be found in \cite {sigma}.)
Numerically,
\begin {equation}
   a_1 = 0.323833\cdots \, .
\label {eq:a1}
\end {equation}

As noted above, the corresponding calculation in Theory 2,
when dimensionally regularized, is trivial, and matching Theory 2 to
Theory 1 merely means setting the collision operator equal to
the result of (\ref {eq:dc2}) and (\ref {eq:Msqr1}), with the same
matrix element $\gamma_1$ appearing in (\ref {eq:gamma}).
This is the bare value of $\gamma_1$ in Theory 2;
to obtain the renormalized value in minimal subtraction,
simply drop the $1/\eps$ pole.

It should be emphasized that our argument that the color conductivity $\sigma$
can be unambiguously defined at NLLO as the parameter of an effective theory
does not apply to $\gamma_1$.
Because Theory 2 is not ultraviolet finite, its bare parameters depend
explicitly on the renormalization scale $\mu$, as seen explicitly in
the above result for $\gamma_1(\mu)$.
This is completely analogous to, for example, the minimal-subtraction
definition of quark mass in QCD.
But a precise, convention independent definition of $\gamma_1$ is irrelevant
(and unnecessary) for the correct evaluation of physical quantities.%
\footnote
    {
    The matrix element $\gamma_1$ differs by a finite regularization
    independent amount from the hard gauge boson damping rate
    \cite {sigma,USA}.
    Many attempts have been made to formulate a gauge-invariant
    ``pole mass'' definition of the damping rate for non-Abelian
    theories, but these have so far led to a mire of confusion \cite{GammaMire}.
    For discussion of the related (but quite distinct) issue in QED,
    see Ref.\ \cite{GammaQED}.
    }


\subsection {Matching Theory 2 to Theory 3}

Matching Theory 2 to B\"odeker's effective theory at NLLO requires a
one-loop calculation in both these theories.
In Ref.\ \cite {sigma}, we describe how
appropriate ratios of large time-like Wilson loops may be chosen as
the physical quantities whose expectations are matched in order
to determine the parameter $\sigma$ of B\"odeker's effective theory.
In practice, this is completely equivalent to
matching the $\omega \ll k$, $k \to 0$ behavior of the $A_0$
self-energy $\Pi_{00}(\omega,k)$ in Coulomb gauge.
We find,
\begin {equation}
   \sigma = {m^2\over d\gamma_1}
       \left\{ 1
          - {\CA \alpha T \over \gamma} \left[
                 {1\over\eps} + \ln\!\left(\bar\mu\over\gamma\right) + b \,
             \right]
          + O\!\left({1 \over \ln^2(1/g)}\right)
       \right\} ,
\label{eq:fsigma}
\end {equation}
where $\bar\mu$ is again the \MSbar\ scale, and the constant $b$ is given by
\begin {equation}
   b = {2\over\pi} \int_0^{\infty} d\rho \left[
      {1\over 2} \left( \Sigma_0(\rho) - {3\over\rho^2}\right)
      + {8\over3\rho^2} \Bigl(1 - \Sigma_1(\rho)\Bigr)
      + {4\over 15} \, \Sigma_2(\rho)
      - {\pi \rho \over 2 (\rho^2+1)}
   \right] .
\label {eq:I}
\end {equation}
Here, $\rho $ represents the dimensionless magnitude $|\p|/T$ of a loop momentum,
and $\Sigma_{m}(\rho)$ are the dimensionless functions defined in bra-ket
notation as
\begin {equation}
   \Sigma_{m}(\rho) =
   \langle mm | (i v_z \rho + \deltaC/\gamma)^{-1} | mm \rangle ,
\end {equation}
where $\deltaC$ is the linear operator defined by (\ref{eq:deltaC}),
the inverse is a non-trivial operator inverse
(because $\deltaC$ and $v_z$ do not commute),
and $|lm\rangle$ represents the function $Y_{lm}(\v)$.
We give formulas for evaluating these functions numerically in
Ref.\ \cite {sigma}.
The numerical result for $b$ is
\begin {equation}
   b = 2.8380\cdots .
\label {eq:b}
\end {equation}


\subsection {Final result}

Combining Eqs.\ (\ref{eq:gamma}) and (\ref{eq:fsigma}) yields
our NLLO result for the color conductivity.
The structure is clearest if we write an expansion for $\sigma^{-1}$
(the ``color resistivity'')
rather than $\sigma$ directly.
One finds
\begin {mathletters}
\label{eq:final}
\begin {eqnarray}
   \sigma^{-1}
   &=& {3\CA \alpha T\over m^2}
       \left[\,
       \ln\!\left(m\over\gamma(\mu)\right)
          + \Const
	  + O\!\left(1 \over \ln(1/g)\right)
       \right] ,
\end {eqnarray}
with
\begin {eqnarray}
    \Const &\equiv&
          \half \ln {\pi\over4}
          + a_1
          + b
    =
          3.0410\cdots \,.
\end {eqnarray}
\label {eq:final2}%
\end {mathletters}%
Note that the $1/\eps$ divergences have canceled, as they must.
Inside the logarithm of (\ref {eq:final}),
$\gamma(\mu)$ is to be understood as simply the leading-log formula
\begin {equation}
   \gamma(\mu) \approx \CA \alpha T \ln\!\left(m\over\mu\right) ,
\end {equation}
and $\mu$ should be chosen so that it is of order $\gamma$.
One may easily verify that the $\mu$ dependence in the NLLO result
(\ref{eq:final}) only affects that
answer at order $[\ln(m/\gamma)]^{-1} \sim [\ln(1/g)]^{-1}$, which is beyond
the order of this calculation.

In order to use the NLLO result (\ref {eq:final}) in any practical calculation
one must, of course, choose some particular value of $\mu$
[and ignore the unknown $O(1/\ln g^{-1})$ corrections].
In the absence of a full next-to-next-to-leading-log analysis,
there is no clearly preferred procedure for determining an ``optimal'' value of $\mu$.
However, one reasonably natural choice is the ``fastest apparent convergence''
(FAC) scheme%
\footnote
    {%
    At our next-to-leading-log order,
    one cannot use any minimal sensitivity criterion,
    since the NLLO result (\ref {eq:final}) has no stationary point in $\mu$.
    }
which is to choose $\mu$ so that the
next-to-leading order correction vanishes.
In the present context, this amounts to choosing the scale $\mu_\FAC$
which satisfies
\begin {equation}
    \mu_{\FAC} = e^{-\Const} \, \gamma(\mu_{\FAC}) \,.
\end {equation}
For this choice, the NLLO conductivity is simply
\begin {equation}
   \sigma^{-1}_\NLLO
   = {3\CA \alpha T\over m^2} \,
       \ln\!\left(m\over\mu_{\FAC}\right) \,.
\end {equation}


\section {Conclusion}

We have shown that at next-to-leading log order, no modification is required
to B\"odeker's effective theory for non-perturbative color dynamics other than
the insertion of the correct NLLO value of the color conductivity.
Using our NLLO result (\ref{eq:final}), one may instantly
generalize Moore's numerical result (\ref{eq:GammaB})
for the topological transition rate of hot electroweak theory to NLLO:
\begin {equation}
   \Gamma \approx (10.8 \pm 0.7)
       \left( {g T \over m} \right)^2
       \alphaw^5 \, T^4
       \left[\,
       \ln\!\left(m\over\gamma(\mu)\right)
          + 3.041
	  + O\!\left(1 \over \ln(1/g)\right)
       \right] .
\label {eq:Gam}
\end {equation}

\begin {figure}[h]
   \begin {center}
      \leavevmode
      
      \begin {picture}(0,0)
	\put(120,-5){$\gamma/\mu$}
	\put(-45,80){$\displaystyle\ln
                \left({e^\Const m \over \gamma(\mu)}\right)$}
	\put(181,22){\LARGE $\downarrow$}
	\put(181,88){\LARGE $\uparrow$}
      \end {picture}
      \epsfbox {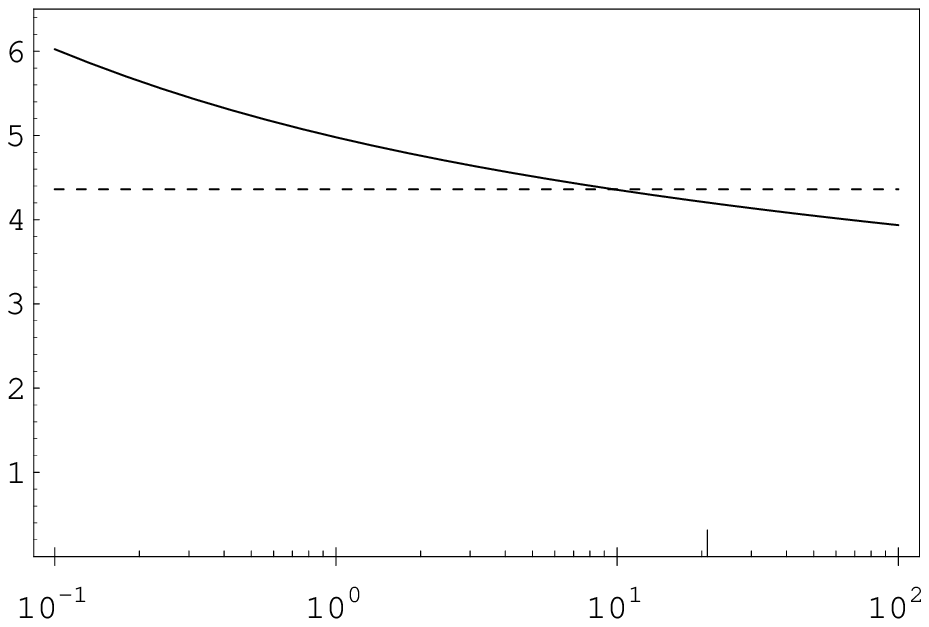}
      \bigskip
   \end {center}
   \caption
	{%
	The value of the $\ln \!\left[m / \gamma(\mu)\right]\!{+}\Const$
        factor,
	appearing in the inverse color conductivity
	(\protect\ref {eq:final})
	and in the topological transition rate (\protect\ref {eq:Gam}),
	plotted as a function of $\gamma/\mu$,
	for electroweak theory with a single Higgs doublet and $g^2 = 0.4$.
	The dashed line indicates the value of $4.4$ for which
	the NLLO result for the topological transition rate
	(\protect\ref {eq:Gam}) (with the unknown yet-higher-order
	$1/\ln g^{-1}$ terms neglected) agrees with independently determined
	results from more microscopic numerical simulations
	\protect\cite {top-trans,moore-SEWM}.
	This value should be regarded as having a significant systematic
	uncertainty of perhaps 20\%.
	The arrow on the abscissa indicates the FAC point where
	$\gamma(\mu)/\mu = e^\Const \simeq 20.93$.
	\label {fig}
	}
\end {figure}

Moore, and others, have also obtained numerical results for the topological
transition rate by using a more microscopic theory (analogous to a lattice
version of what we called Theory 1) \cite {top-trans,moore-SEWM}.
Moore also attempted to estimate the size of the NLLO correction to $\Gamma$
by fitting the results of these simulations to the functional form
\begin {equation}
    \Gamma = \kappa \left({gT \over m}\right)^2 \alphaw^5 \, T^4
    \left[\,
       \ln\!\left(m\over g^2 T\right)
          + \delta \,
       \right] ,
\label {eq:guess}
\end {equation}
with the value of $\kappa$ fixed to 10.8, as determined from simulations
of B\"odeker's effective theory.
This led to $\delta \approx 3.6$, with perhaps 20\% uncertainty due to
systematic errors \cite {moore-SEWM}.%
\footnote
    {%
    Lattice artifacts exist in these more microscopic simulations
    (due, in particular, to the lattice dispersion relation allowing
    unwanted Cherenkov radiation),
    which cause them not to reproduce, precisely, the dynamics
    responsible for NLLO corrections to $\Gamma$.
    These effects have only been crudely estimated, but are
    a major part of the uncertainty in the estimate of $\delta$.
    }
This implies an estimate of $4.4$ for the value of the
$\ln (m / g^2 T) {+} \delta$
factor in (\ref {eq:guess}), which
may be compared with the square bracket appearing in
(\ref {eq:Gam}).
This comparison is presented graphically in Fig.~\ref{fig}.
The solid line is a plot of
$\ln \!\left[m / \gamma(\mu)\right]\!{+}\Const$ as a function of
$\gamma/\mu$, for the specific case of electroweak theory
with a single Higgs doublet and $g^2 = 0.4$.
The dashed line indicates the value of $4.4$ estimated in \cite {moore-SEWM}.
The arrow on the abscissa indicates the FAC point, where
$\gamma(\mu)/\mu = e^\Const \simeq 20.93$.

The similarity between our NLLO result
and the value
inferred from numerical simulations is remarkable.
There was no obvious reason {\it a priori}\/ why it should be a
reasonable approximation to treat logarithms of the gauge coupling
[that is, $\ln(\#/g)$],
as large for physical
values of the coupling.  However, one sees from
Fig.\ \ref{fig} that even the most naive
prescription of picking $\mu$ exactly equal to the scale $\gamma$
produces a result within 20\% of Moore's numerical estimate (which itself
has an uncertainty estimated at 20\%).
The naive prescription of letting $\mu$ vary by an entire order of
magnitude about $\gamma$ (from $0.1\gamma$ to $10\gamma$), and taking
the variation as a crude estimate of the uncertainty from leaving
out higher-order corrections, still keeps the result within roughly 35\%.
The FAC value of $\gamma(\mu)/\mu$ suggests that perhaps one should pay
more attention to the range $\mu < \gamma$, where the variation with $\mu$ is
even smaller and the agreement with Moore's numerical estimate even
better.

The close agreement between the FAC value of $\mu$ and the precise
point where the curves of Fig.\ \ref{fig}
cross is striking, but probably fortuitous
given the uncertainty in the numerical simulation value.
Nevertheless, perhaps the characteristic size of
neglected corrections really is $e^{-\Const} \gamma$,
and not just $\gamma$ (as one might naively expect).
If true, this would mean that (for electroweak theory)
the expansion in inverse logs actually has a respectably small
expansion parameter of about 0.25.
%
To confirm (or refute) this expectation, one would need
to calculate to yet higher order in the expansion in inverse logarithms.
As discussed earlier, this presumably
requires adding new operators to B\"odeker's
effective theory, and this in turn implies that
new numerical simulations would be required.
Such additional higher-dimension operators will also cause
the effective theory to no longer be UV finite, implying that a careful
matching calculation for the lattice-regularized theory would now be required.

\section* {ACKNOWLEDGMENTS}

We thank Guy Moore and Dietrich B\"odeker
for a variety of helpful discussions.
This work was supported, in part, by the U.S. Department
of Energy under Grant Nos.~DE-FG03-96ER40956
and DE-FG02-97ER41027.

\clearpage
\appendix

\section{Power counting with functional integrals}

In the main text, we discussed how to count the relative scaling dimensions of
operators appearing in the Langevin equation.
We swept under the rug issues of how
to do the same for terms involving the noise.  To be systematic, it is
much more convenient to recast the problem in terms of a functional integral,
where dimension counting of operators is more familiar.  Also, in the
functional integral version, 
noise has been integrated over, and
need not be discussed separately.

The functional integral representation of B\"odeker's theory is well known
as the functional integral representing stochastically quantized 3-dimensional
non-Abelian gauge theory.  See, for example, Ref.\ \cite{Zinn-Justin}
for a review, or \cite{thy2} for a discussion in the present context.
In $A_0=0$ gauge, one form is
\begin {mathletters}
\begin {equation}
   Z = \int [{\cal D}\A(\x,t)] \> J[\A] \> e^{-S[\A]}
\end {equation}
with
\begin {equation}
   S = {1\over 4\sigma T} \int dt \> d^3x \>
          \left| \, \sigma \, {d\A\over dt} + \D\times\B \right|^2
   ,
\label {eq:SA0}
\end {equation}%
\label {eq:SA}%
\end {mathletters}%
and where $J[\A]$ is a
Jacobian factor that equals one in dimensional regularization.
It is useful to rewrite the action in terms of the magnetic energy
density,
\begin {equation}
   \V(\A) \equiv \half \B^a \cdot \B^a ,
\end {equation}
as
\begin {equation}
   S = {1\over 4\sigma T} \int dt \> d^3x \>
          \left| \, \sigma \, {d\A\over dt} + {d\V\over d\A} \right|^2
   .
\end {equation}

There is a hidden supersymmetry of this functional integral which restricts
the form of interactions which can be added to it.  The functional integral
can be rewritten in superfield notation as
\begin {equation}
   Z = \int [{\cal D}\bPhi] \> e^{-S[\bPhi]} ,
\label{eq:Zsusy}
\end {equation}
with
\begin {equation}
   S = {1\over T} \int d\theta \> d\bar\theta \> dt \> d^3x \>
       \left[\,
          \sigma \, \bar {\cal D}_t \bPhi \cdot {\cal D}_t \bPhi
          + \V(\bPhi)
       \right] ,
\label{eq:Ssusy}
\end {equation}
where $\bPhi$ is the superfield
\begin {equation}
   \bPhi = \A + \bar\theta\c + \bar\c\theta + \blambda\bar\theta\theta .
\label {eq:Phi}
\end {equation}
The supersymmetry only involves time, rather than space-time, and the
SUSY time derivatives are
\begin {equation}
   \bar {\cal D}_t \equiv \partial_\theta \,,
   \qquad\qquad
   {\cal D}_t \equiv \partial_{\bar\theta} - \theta \partial_t \,.
\end {equation}
In (\ref {eq:Phi}),
$\blambda$ is an auxiliary field, and $\c$ and $\bar\c$ are ghost fields,
the integral over which produces the Jacobian factor $J[\A]$ mentioned
earlier.

The scaling dimensions of all the fields may now easily be read off
from the supersymmetric action:
$[t] = [x]^2$,
$[\theta]=[\bar\theta]=[x]$,
and $[\bPhi] = [x]^{-1/2}$.
(Keep in mind that for Grassmann integration,
$d\theta$ has the inverse dimension
of $\theta$, and so $[d\theta] = [d\bar\theta] = [x]^{-1}$.)
One may now analyze the dimensions of what possible irrelevant interactions
can be added to the supersymmetric action (\ref{eq:Ssusy}) consistent
with gauge invariance, parity, and supersymmetry.%
\footnote
    {
    Although it is not required for our discussion,
    it is worth noting that time-reversal invariance also
    constrains the possible irrelevant interactions which
    may appear in the effective theory.
    This may be surprising at first sight,
    since the Langevin equation
    (\ref {eq:3}) defining the leading-order effective theory
    is dissipative and manifestly violates time reversal invariance.
    Nevertheless, this theory generates time-dependent equilibrium
    correlation functions which do respect time-reversal symmetry.
    Formally, this is easiest to see from the functional integral
    (\ref {eq:SA}).
    The action density appearing in (\ref {eq:SA0}) is not invariant
    under time reversal.
    However, after multiplying out the square, it is only the
    cross term which is time reversal odd.
    And this cross term is a total time derivative,
    $
	\int d^3x \> \sigma \, (d\A/dt) \cdot \D \times \B
	=
	(d / dt) \int d^3 x \> \sigma \, {\cal V}(\A)
    $.
    Hence, the action (\ref {eq:SA}) is in fact time-reversal invariant 
    up to boundary terms at $t = \pm\infty$, which are irrelevant
    as far as equilibrium properties (including time-dependent
    correlation functions) are concerned.
    This invariance (up to boundary terms) must remain true in the
    presence of higher-dimension irrelevant interactions.
    For related discussion in terms of the Fokker-Planck equation,
    see Ref.~\cite {Arnold-langevin}.
    }
This provides a more familiar way to do the power counting analysis
than that of the main text.
For instance, consider the interaction
$\B D^2 \B$, where $\B$ and $D$ are to be understood as the normal
non-SUSY expressions for the magnetic field and the gauge-covariant
derivative, but with $\A$ replaced by the superfield $\bPhi$.
The interaction $\B D^2 \B$ has
scaling dimension $[x]^{-5}$, as opposed to the other terms in
the supersymmetric Lagrangian, which have the marginal scaling
$[x]^{-3}$.  So the effects of this term should be suppressed by
at least $(k/\gamma)^2$, since $\gamma$ serves as the cut-off scale $\Lambda$
for this theory.
Another example is
$\bar {\cal D}_t \bPhi \cdot {\cal D}_t \bPhi \times \B$
(still keeping to $A_0=0$ gauge),
which has scaling dimension $[x]^{-9/2}$, and so is suppressed by
$(k/\gamma)^{3/2}$.
In fact, the coefficient of this operator must also contain a factor
of $g$, because the underlying theory is unchanged if
$g \to -g$ and $\A \to -\A$ (or $\bPhi \to -\bPhi$).
The effective theory must also respect this invariance.
However a suitable rescaling of variables in Theory 2
(see Ref.~\cite {sigma})
shows that the theory only depends on the
scale $\gamma$ plus
the dimensionless combination $g^2T / \gamma$.
Hence, a factor of $g$ in the coefficient of an induced higher dimension
operator implies that the effects of this operator will have
an additional suppression by $(g^2 T/\gamma)^{1/2}$.
Therefore, the effects of the interaction
$\bar {\cal D}_t \bPhi \cdot {\cal D}_t \bPhi \times \B$
must be suppressed by $(g^2 T/\gamma)^{1/2} \, (k/\gamma)^{3/2}$.

A little thought shows that the net result is that the only interactions
which can be added to the action (\ref{eq:Ssusy}),
consistent with its symmetries,
are either (i) irrelevant operators whose effects, at the scale $g^2 T$
are suppressed by at least two powers of $g^2 T/\gamma$,
or (ii) total derivatives [such as
$({\cal D}_t\bPhi) \cdot \D \times \B = {\cal D}_t {\cal V}(\bPhi)$]
which have no effect on the dynamics.




\begin {references}

\bibitem{baryogenesis}
  For reviews, see
   A. Cohen, D. Kaplan and A. Nelson,
   {\sl Annu.\ Rev.\ Nucl.\ Part.\ Sci.}\ {\bf 43}, 27 (1993);
   V. Rubakov and M. Shaposhnikov,
   {\tt hep-ph/9603208},
   {\sl Usp.\ Fiz.\ Nauk} {\bf 166}, 493 (1996)
   [{\sl Phys.\ Usp.}\ {\bf 39}, 461 (1996)];
   M. Trodden,
   {\tt hep-ph/9803479},
   Case Western report no.\ CWRU-P6-98.

\bibitem {bodeker}
    D. B\"odeker,
    {\tt hep-ph/9810430},
    {\sl Phys.\ Lett.}\ {\bf B426}, 351 (1998);
    {\tt hep-ph/9905239};
    {\tt hep-ph/9903478}.

\bibitem{moore}
   G. Moore,
   {\tt hep-ph/9810313}.

\bibitem {notransition}
   K. Kajantie, M. Laine, K. Rummukainen, and M. Shaposhnikov,
   {\tt hep-ph/9605288},
   {\sl Phys.\ Rev.\ Lett.}\ {\bf 77}, 2887 (1996).
   S. Elitzur,
   {\sl Phys.\ Rev.\ D} {\bf 12}, 3978 (1975).

\bibitem {Blog1}
   P. Arnold, D. Son, and L. Yaffe,
   {\tt hep-ph/9810216},
   {\sl Phys.\ Rev.\ D} {\bf 59}, 105020 (1999);
   {\tt hep-ph/9901304},
   {\sl Phys.\ Rev.\ D} {\bf 60}, 025007 (1999).

\bibitem{scale}
    T. Appelquist and R. Pisarski,
      {\sl Phys.\ Rev.\ D} {\bf 23}, 2305 (1981);
    S. Nadkarni,
      {\sl Phys.\ Rev.\ D} {\bf 27}, 917 (1983);
      {\sl Phys.\ Rev.\ D} {\bf 38}, 3287 (1988);
      {\sl Phys.\ Rev.\ Lett.}\ {\bf 60}, 491 (1988);
    N. Landsman,
      {\sl Nucl.\ Phys.}\ {\bf B322}, 498 (1989);
    K. Farakos, K. Kajantie, M. Shaposhnikov,
      {\sl Nucl.\ Phys.}\ {\bf B425}, 67 (1994).

\bibitem{Selikhov&Gyulassy}
    A. Selikhov and M. Gyulassy,
    {\tt nucl-ph/9307007},
    {\sl Phys.\ Lett.}\ {\bf B316}, 373 (1993).

\bibitem {alpha5}
   P. Arnold, D. Son, and L. Yaffe,
   {\tt hep-ph/9609481},
   {\sl Phys.\ Rev.\ D} {\bf 55}, 6264 (1997).
   See also
   P. Huet and D. Son,
   {\tt hep-ph/9610259},
   {\sl Phys.\ Lett.}\ {\bf B393}, 94 (1997);
   P. Arnold,
   {\tt hep-ph/9701393},
   {\sl Phys.\ Rev.\ D} {\bf 55}, 7781 (1997).

\bibitem{classical}
   D. Grigorev and V. Rubakov,
   {\sl Nucl.\ Phys.}\ {\bf B299}, 67 (1988);
   D. Grigorev, V. Rubakov, and M. Shaposhnikov,
   {\sl Nucl.\ Phys.}\ {\bf B326}, 737 (1989);
   for a review, see
   E. Iancu,
   {\tt hep-ph/9807299},
   Saclay report no.\ SACLAY-T98-073.

\bibitem {flow gauges}
   H. Chan and M. Halpern,
   {\sl Phys.\ Rev.\ D} {\bf 33}, 540 (1985).

\bibitem {zinnjustin&zwanziger}
   J. Zinn-Justin and D. Zwanziger,
   {\sl Nucl.\ Phys.}\ {\bf B295} [FS21], 297 (1988).

\bibitem {thy2}
    P.~Arnold,
    {\it An effective theory for $\omega \ll k \ll gT$ color dynamics
    in hot non-Abelian plasmas},
    {\tt hep-ph/9912307}.

\bibitem {sigma}
    P.~Arnold and L.~Yaffe,
    ``{\it High temperature color conductivity at next-to-leading log order},''
    {\tt hep-ph/9912306}.

\bibitem{stochastic}
   For a review, see
   P. Damgaard and H. Huffel,
   {\sl Phys.\ Rept.}\ {\bf 152}, 227 (1987).

\bibitem{gammag}
   R. Pisarski,
   {\sl Phys.\ Rev.\ Lett.}\ {\bf 63}, 1129 (1989);
   {\sl Phys.\ Rev.\ D} {\bf 47}, 5589 (1993).

\bibitem{manuel}
   D. Litim and C. Manuel,
   {\tt hep-ph/9902430},
   {\sl Phys.\ Rev.\ Lett.}\ {\bf 82}, 4981 (1999).

\bibitem{basagoiti}
   M. Basagoiti,
   {\tt hep-ph/9903462}.

\bibitem {b&i}
    J.-P.~Blaizot and E.~Iancu,
    {\tt hep-ph/9903389},
    {\sl Nucl.\ Phys.}\ {\bf B557}, 183 (1999).

\bibitem {HTL}
    E. Braaten and R. Pisarski,
    {\sl Phys.\ Rev.\ D} {\bf 45}, 1827 (1992);
    {\sl Nucl.\ Phys.}\ {\bf B337}, 569 (1990);
    S. Mr\'{o}wczy\'{n}ski,
    {\sl Phys.\ Rev.\ D} {\bf 39}, 1940 (1989);
    H.-Th. Elze and U. Heinz,
    {\sl Phys.\ Rept.}\ {\bf 183}, 81 (1989);
    J. Blaizot and E. Iancu,
    {\sl Nucl.\ Phys.}\ {\bf B417}, 609 (1994);
    and references therein.

\bibitem {iancu}
    E.~Iancu,
    {\tt hep-ph/9710543},
    {\sl Phys.\ Lett.}\ {\bf B435}, 152 (1998).

\bibitem{jj}
   P. Arnold and L. Yaffe,
   {\tt hep-ph/9709449},
   {\sl Phys.\ Rev.\ D} {\bf 57}, 1178 (1998).

\bibitem {viscosity}
    H.\ Heiselberg,
    {\tt hep-ph/9401309},
    {\sl Phys.\ Rev.\ D} {\bf 49}, 4739 (1994).

\bibitem{Lebedev&Smilga}
   V. Lebedev and A. Smilga,
   {\sl Phys.\ Lett}. {\bf 253B}, 231 (1991);
   {\sl Ann.\ Phys.}\ (NY) {\bf 202}, 229 (1990);
   {\sl Physica} {\bf A181}, 187 (1992).

\bibitem {USA}
   J. Blaizot and E. Iancu,
   {\tt hep-ph/9906485}.

\bibitem{Braaten&Nieto}
    E. Braaten and A. Nieto,
    {\sl Phys.\ Rev.\ D}\ {\bf 51}, 6990 (1995); {\bf 53}, 3421 (1996).

\bibitem{KLRS}
    K. Kajantie, M. Laine, K. Rummukainen, M. Shaposhnikov,
    {\sl Nucl.\ Phys.}\ {\bf B458}, 90 (1996).

\bibitem {constant-field}
    L.~Brown and W. Weisberger,
    {\sl Nucl.\ Phys.}\ {\bf B157}, 285 (1979).

\bibitem {pi}
   H. Weldon,
     {\sl Phys.\ Rev.\ D} {\bf 26}, 1394 (1982);
   U. Heinz,
     {\sl Ann.\ Phys.}\ (N.Y.) {\bf 161}, 48 (1985); {\bf 168}, 148 (1986).

\bibitem {GammaMire}
   S. Peign\'e, E. Pilon, D. Schiff,
   Z. Phys.\ {\bf C60}, 455 (1993);
   R. Baier, H. Nakkagawa, A. Niegawa,
   {\sl Can.\ J. Phys.}\ {\bf 71}, 205 (1993);
   T. Altherr, E. Petitgirard, T. del Rio Gaztellurutia,
   {\sl Phys.\ Rev.\ D} {\bf 47}, 703 (1993);

\bibitem {GammaQED}
   J. Blaizot and E. Iancu,
   {\sl Phys.\ Rev.\ D} {\bf 55}, 973 (1997); {\bf 56}, 7877 (1997);
   {\sl Phys.\ Rev.\ Lett.}\ {\bf 76}, 3080 (1996);
   K. Takashiba,
   {\sl Int.\ J.\ Mod.\ Phys.}\ {\bf A11}, 2309 (1996);
   D. Boyanovsky and J. de Vega,
   {\sl Phys.\ Rev.\ D} {\bf 59}, 105019 (1999).

\bibitem {top-trans}
    G.~Moore, C.~Hu, B.~Muller, {\tt hep-ph/9710436},
    {\sl Phys.\ Rev.\ D} {\bf 58}, 045001 (1998);
    G.~Moore and K.~Rummukainen, {\tt hep-ph/9906259};
    D.~B\"odeker, G.~Moore and K.~Rummukainen,
    {\tt hep-ph/9907545},
    {\tt hep-lat/9909054}.

\bibitem {moore-SEWM}
    G.~Moore, {\tt hep-ph/9902464}.

\bibitem {Zinn-Justin}
    J.~Zinn-Justin, {\sl Quantum Field Theory and Critical Phenomena},
    2nd edition (Oxford University Press, 1993).

\bibitem {Arnold-langevin}
    P. Arnold,
    {\it ``Langevin equations with multiplicative noise: resolution of
      time discretization ambiguities for equilibrium systems,''}
    {\tt hep-ph/9912208}.

\end {references}
\end {document}